\documentclass[12pt]{article}

\usepackage[margin=1in]{geometry}
\usepackage{setspace}
\usepackage[colorlinks=true, linkcolor=black, citecolor=black, urlcolor=black]{hyperref}
\usepackage{titlesec}
\usepackage{booktabs}
\usepackage{longtable}
\usepackage{natbib}  
\usepackage{indentfirst}
\usepackage{amssymb}
\begin{document}

\begin{center}
  {\LARGE\bfseries Levers of Power in the Field of AI}\\[0.3em]
  {\normalsize An Ethnography of Personal Influence in Institutionalization}
\end{center}

\vspace{0.8em}

\begin{center}
  \begin{minipage}[t]{0.25\textwidth}
    \centering
    {\footnotesize \textbf{Tammy Mackenzie}}\\
    {\footnotesize The Aula Fellowship, Montreal, Canada}\\
    {\footnotesize \texttt{tammy@theaulafellowship.org}}
  \end{minipage}
  \hspace{0.05\textwidth}
  \begin{minipage}[t]{0.25\textwidth}
    \centering
    {\footnotesize \textbf{Sukriti Punj}}\\
    {\footnotesize The Aula Fellowship, Delhi, India}\\
    {\footnotesize \texttt{sukriti@theaulafellowship.org}}
  \end{minipage}
  \hspace{0.05\textwidth}
  \begin{minipage}[t]{0.25\textwidth}
    \centering
    {\footnotesize \textbf{Natalie Perez}}\\
    {\footnotesize The Aula Fellowship, Honolulu, United States}\\
    {\footnotesize \texttt{neichner@hawaii.edu}}
  \end{minipage}
\end{center}

\vspace{0.9em}

\begin{center}
  \begin{minipage}[t]{0.45\textwidth}
    \centering
    {\footnotesize \textbf{Sreyoshi Bhaduri}}\\
    {\footnotesize The Aula Fellowship}\\
    {\footnotesize New York, United States}\\
    {\footnotesize \texttt{sreyoshibhaduri@gmail.com}}
  \end{minipage}
  \hspace{0.05\textwidth}
  \begin{minipage}[t]{0.45\textwidth}
    \centering
    {\footnotesize \textbf{Branislav Radeljić}}\\
    {\footnotesize The Aula Fellowship}\\
    {\footnotesize London, United Kingdom}\\
    {\footnotesize \texttt{branislav@theaulafellowship.org}}
  \end{minipage}
\end{center}

\vspace{0.9em}

\vspace{-0.6em} 

\renewcommand{\abstractname}{\large\bfseries Abstract}
\begin{abstract}
\normalsize
\noindent
This paper examines how decision makers in academia, government, business, and civil society navigate questions of power in implementations of artificial intelligence (AI). The study explores how individuals experience and exercise “levers of power,” which are presented as social mechanisms that shape institutional responses to technological change. The study reports on the responses of personalized questionnaires designed to gather insight on a decision maker’s institutional purview, based on an institutional governance framework developed from the work of Neo Institutionalists. Findings present the anonymized, real responses and circumstances of respondents in the form of twelve fictional personas of high-level decision makers from North America and Europe. These personas illustrate how personal agency, organizational logics, and institutional infrastructures may intersect in the governance of AI. The decision makers’ responses to the questionnaires then inform a discussion of the field level personal power of decision-makers, methods of fostering institutional stability in times of change, and methods of influencing institutional change in the field of AI. The final section of the discussion presents a table of the dynamics of the levers of power in the field of AI for change makers and 5 testable hypotheses for institutional and social movement researchers. In summary, this study provides insight on the means for policymakers within institutions and their counterparts in civil society to personally engage with AI governance.
\end{abstract}

\vspace{4pt}
\noindent\textit{\textbf{Keywords:} artificial intelligence, power, institutionalism, policymaking, civil society advocacy, governance}


\section{Introduction}

This short and preliminary report on the results of the questionnaires informs four contributions to scholarship on social change and individual power in the field of AI, specifically within constituent institutions and large organizations and the larger social environment in which they operate. Presented here are discussions of the relative personal power of respondents, on fostering institutional stability, and on the work of influencing institutional change. The discussion concludes with a set of testable hypotheses about the dynamics of the "levers of power" in this field.

This work proceeds from the assumption that decision makers in all parts of society have decisions to make about AI, and furthermore from the statements by organizations such as Mila Quebec Artificial Intelligence Institute, who have called for massive engagement from civil society, on behalf of regulators and technology companies \citep{Mila,Kandikatla2025,Khan2025,Mackenzie2024,Mackenzie2025}
It may be reasonable to generalize that most of the people involved in doing AI regulation recognize this need for engagement from society, whether from a logic of democratic representation or from a commercial logic of matching the product to the demand. 

When we consider the dynamics of institutionalization of social practices, “levers of power” are the means by which individuals can use their personal agency to affect how an institution responds to events. In the questionnaires, respondents were asked about the means they have to affect our institutions and the extent to which“levers of power” can be exerted upon them by others, in their capacity as decision makers within their respective institutions. The purpose of this study is to inform an understanding of the power dynamics in the field of AI, and to inform the design of a set of dynamic control indicators and contingency mechanisms for people responding to existing and upcoming AI situations.

This research builds on the research-for-impact process \citep{New2020}. It aims to give policymakers and civil society advocates a clearer view of how different decision makers approach AI within their own domains. By understanding these perspectives, people can design strategies and programs that fit the specific context of each institution. The goal is to strengthen individual participation in shaping institutions, from positions within an institution and from society at large. For example, an elected representative can learn different means of negotiating power from the senator or from the bureaucrat. A civil society advocate or an academic can learn the same about the channels of power, and use that information to share crucial information about context with the people who can act upon it. 

We used an ethnographic research design with three core components: (1) basic demographic questions about the respondent's position in society, (2) Likert-scale questions about their views of different power dynamics, and (3) open-ended questions about their views and lived experiences. Specifically, the questionnaires were designed around a modified framework of the dimensions of institutional governance, developed within the traditions of Social Institutionalism by \citet{Hinings2017}, and extended for consideration of the field of AI with an additional dimension, \textit{Idea Mobility}, as per Scandinavian Institutionalism \citep{Olsen2009}. Collectively, the dimensions of the modified framework are referred to herein as levers of power.

The modified framework provides a list of the ways in which power can be enacted within an ``institutional field,'' also sometimes called ``organisational field,'' understood here to be a system which encompasses all the practices of organizations and individuals to do with a given topic \citep{Powell1983}. In consideration of AI in society, the institutional field can be usefully categorized into broad sectors, such as academia, civil society, government, and business. Within the field there are different means by which people can collaborate to bring about changes in practice. These means, called levers of power, range from formal and systemized to less formal and more diffuse in society. 

There are many different levers and different ways of naming them. In the modified framework that we have applied herein, the levers of power considered are: logics, relative elaboration of the institutional infrastructure, governance, collective interest organizations, regulators, informal governance bodies, field configuring events, status differentiators, organizational models or templates, categories or labels, norms, relational channels, and idea mobility.

Questions were prepared to provide insights on each lever of power. There were four basic models of questionnaire prepared, one each for academia, civil society, business, and government. These were then further customized for the respondent based on their specific questions and institutional role(s). These were first tested by an academic and a business respondent, both of whom provided detailed comments. Per \citet{Malmqvist2019}, these steps were used to pilot test the questionnaire before launching it to the respondents, primarily to ensure the quality and clarity of the questions. Following these first responses and comments, some questions were clarified and explanations for technical terms were added. Respondents were also asked if the questionnaire took the expected amount of time (15 minutes). Those who responded to that question said that it had. One sample questionnaire is provided in the annex (academic version). The research team has collected 53 responses from each decision maker who joined the study. 

This study included 12 participants who are key decision makers in their respective fields. Most respondents to the personalized questionnaire opted to remain anonymous, and so we opted to anonymize them all. All respondents in this study met the sampling criteria which requires the respondents to hold considerable power in determining institutional strategy for large groups of people, including in government, in academia, in civil society, and in business. Specifically, all have considerable power in determining how AI is being implemented in the different spheres of our society. Questionnaire responses were cross-referenced between respondents, considered in light of the levers of power and further contextualized by a documentary examination of the respondent's professional body of work. We celebrate the idiographic with this study. That is, a nomothetic viewpoint, which is common in science fields, seeks to generalize and determine broader understandings of the world. On the other hand, an idiographic perspective tends to focus on deeper understandings of specific human experiences. 

The authors of the present study are scholars who work across sectors, including in academia, in industry, and in policymaking. They are Fellows of The Aula Fellowship, a non-profit think tank with the mission to ensure that everyone can access the conversation on AI. To that end, this study is meant to provide a bridge between communities and the decision makers within our institutions. 

In the following sections, we present personas, which are anonymized portraits of the individual respondents, using their actual responses. This is followed by a discussion of their personal power, of the means by which institutional changes may be accomplished, and presents several testable hypotheses on the dynamics of institutional change, from the point of view of people who are seeking leverage with which to steer such changes.

\section{Personas}

These are anonymized portraits of decision makers in society, resulting from their real responses to personalized questionnaires on how they operationalize their power in society, and how this may be affected by AI. Only the socio-cultural aspects have been anonymized. The quotes and specifics are real.

{
\setlength{\parindent}{0pt}
\setlength{\parskip}{0pt}

\subsection*{Alex}
\textbf{Position:} Policymaker \\
\textbf{Sector:} Academia \\
\textbf{Field:} AI Research and Policy \\[4pt]
Alex works in an established AI research institute, having a mixed organizational structure (hierarchical and horizontal) and interdependent functions. Established and historical practices have a minimal role in their office. They state that they have minimal influence in their organization and have an informal level of interaction with superiors and subordinates. Alex regularly interacts with “non-profit organizations, academic orgs and foundations.” Their job is to “generate policy options with input from specialists, policymakers, and impacted people.” They identified a major (international) policy conference as a field-configuring event on AI, noting a “breadth of expertise, policy-science interface conversations and diplomatic/geopolitical tensions.”

\subsection*{Bailey}
\textbf{Position:} Researcher \\
\textbf{Sector:} Academia \\
\textbf{Field:} Political Science \\[4pt]
Bailey works in a moderately mature university where the significance of established practices is minimal but not non-existent. With two decades of experience, their organization is structurally hierarchical and has rules and norms that are sometimes changed. When asked about an influential event in their professional history, they related: “It was a doctoral workshop to which I was invited to contribute. It was a small group, and in preparation for the workshop (2 days), everybody had to read everybody else's papers and comment. This was extremely productive and beneficial.”

\subsection*{Cameron}
\textbf{Position:} Senator \\
\textbf{Sector:} Government \\
\textbf{Field:} Public Policy, Economy \\[4pt]
A former entrepreneur, Cameron has seven years of experience working in government. The senate has a hierarchical structure, where driving change is a long-drawn process. Established practice has a strong significance in his place of work. He reports that he has little influence over the people he works with. About his interactions with collective interest organizations, Cameron said: “Our job requires us to deal with such an incredibly wide diversity of important issues, with so many conflicting viewpoints, it would be impossible to consider all of the important nuances without their help. Of course, I validate their claims independently. Of course, I ask them to tell me who disagrees with them, and what their issues are, as a way to ultimately verify their integrity. If a major contrary issue emerges that they did not raise, I tell them that their perspective will be immediately discounted.” He continues: “My past life of experience as an entrepreneur is, consistently, what I draw from with every issue.” As for their view of AI in society: “We are not prepared, in any way. Not in terms of threats or opportunities.”

\subsection*{Dana}
\textbf{Position:} Architect in Responsible AI \\
\textbf{Sector:} Business \\
\textbf{Field:} Business, Diligence \\[4pt]
Dana has one year of experience as a developer in a large company designing responsible AI frameworks. They have formal education in computer science, machine learning, and AI ethics (algorithmic fairness and privacy-preserving machine learning algorithms). Their organization is structurally hierarchical and well established. They work in the ombudship service and on the AI and data governance of their organization. Dana intends to solve social problems by creating “literacy programs and defining technical requirements to get it right or minimize the risk.” In their work, they combine “humanitarian aid work with formal education in computer science, machine learning, and AI ethics (algorithmic fairness and privacy-preserving machine learning algorithms).”

\subsection*{Emerson}
\textbf{Position:} Senior Rabbi and Community Leader \\
\textbf{Sector:} Civil Society \\
\textbf{Field:} Religious Organization \\[4pt]
Emerson previously served as a grand chaplain, and their work includes community coordination and spiritual leadership. They have approximately 20 years of experience in this position and interact with social action and charitable groups. Their organization is hierarchical, and their conversations with subordinates are governed by a formal policy “to a certain degree.” Their work is a vocation: “I aspire to bring comfort and consolation to the ill, dying, and the grief-stricken. To help those who are sick, particularly with mental illness.” When asked whether they have concerns about AI in their work, they said: “I have no concerns with AI in my work. Aspects of AI have become invaluable tools for helping others.”

\subsection*{Finley}
\textbf{Position:} Researcher \\
\textbf{Sector:} Academia \\
\textbf{Field:} Science and Education \\[4pt]
Finley works in a highly established and complex organization. Their work at the university depends on established practices. They have been in this position for two years and participate in the elaboration of regulations in their organization indirectly, through a representative. Their reach within the organization is limited to a department-level scope. When asked what their superhero name might be, they replied: “Breaker of Stereotypes!” They share: “I think the biggest concern is not with academics but students. I know of a person (in another country) who wrote their PhD thesis with ChatGPT and managed to get it approved. That is quite scary. I think people should be empowered to use AI to benefit from it but not to the point of managing to get away with breaking the rules.”

\subsection*{Harper}
\textbf{Position:} Executive \\
\textbf{Sector:} Business \\
\textbf{Field:} Data Science \\[4pt]
Harper has four years of experience in their current executive role and is working towards future promotions within the organization. They work in a highly established, hierarchical organization, which has a moderate influence of established practice in its daily work. They are not part of the elaboration of regulations. They consider themselves to be “a builder of products for people, by people.”

\subsection*{Jaime}
\textbf{Position:} Public Servant \\
\textbf{Sector:} Government \\
\textbf{Field:} AI Policy \\[4pt]
Jaime is “a public servant who works in the AI policy space.” They are an experienced public servant and have been in this particular position for several months. They work in a highly established, hierarchical organization and do not participate in the elaboration of regulations. Jaime conducts research and policy analysis for senior civil servants and attends government meetings as an observer. Anonymity is vitally important for Jaime. They have asked: “Please do not attribute information to me that could lead to me being personally identified. I wish to remain anonymous.” They specify why: “I am unable to answer most questions for ethical, political, safety, and security reasons. Please do not attribute any information to myself, my office, or department.”

\subsection*{Kinsey}
\textbf{Position:} Executive Director \\
\textbf{Sector:} Civil Society \\
\textbf{Field:} Technology \\[4pt]
Kinsey is the executive director of a new and emerging think tank on responsible AI. Their organization is structurally horizontal, dynamic, and flexible, with moderate significance of established practice in their work. They have no formal oversight in their position, and their work is accountable to a board, civil society, and governments. They relate an incident indicating the importance of informal norms in the exercise of social power: “My mother-in-law heard I was going to address a committee of the UN. I wondered if I really had anything at all to say. She took me by the shoulders and told me that not only do I have something to say, I am blessed with the ability to say it, so I darn well should. So, I have.” Their superhero name would be: “Doing-the-chores-that-need-doing-person.”

\subsection*{Lindsay} 
\textbf{Position:} Scientific Director \\
\textbf{Sector:} Academia \\
\textbf{Field:} Technology \\[4pt]
Lindsay works as a scientific director of a research institute in a well-established, non-hierarchical, and interdependent organization where rules exist but are flexible and can evolve. Their decision-making is guided by their values and institutional norms. They have more than ten years of experience in this role and several decades as a researcher. They share their principal concern about AI in their sector: “We do not know yet how we would control AI that could be smarter than us, yet we collectively race forward, taking an existential risk for humanity’s future.”
}

\section{Discussion}

This short and preliminary report on the results of the questionnaires informs three contributions to scholarship on social change and individual power in the field of AI, specifically within constituent institutions and large organizations and the larger social environment in which they operate. Presented here are a discussion of the relative personal power of respondents, a discussion of how this informs the work of influencing institutional change, and a set of testable hypotheses about the dynamics of the levers of power in this field. 

\subsection{The relative personal power of decision makers}

Research on imposter syndrome in knowledge work in academia has consistently found that it is a complex social construction, with complex social consequences (Breeze, 2018, Kamran Siddiqui, 2024). In this work, we assume that this holds true generally for other sectors, that the geopolitical and sociotechnical environments in which people find themselves play a variable but significant part in their relative sense of personal empowerment, and that a sense of empowerment precedes the exercise of power. An examination of how people develop that sense of power and of how useful it is to the successful exercise of power are out of the scope of this work, and also not a usual subject in institutionalism. Instead, the present work is on how changes to the institutional field can be encouraged by individuals, and is a question of the interplays between personal agency and structure. From an analysis of responses to the questionnaires, we can discern some common elements between how these decision makers view their personal power. For example, many of the respondents are relatively easy to reach with a message, but few have regular interactions with interest groups outside their institution. For another, though people with more seniority tended to have more of a feeling of power within their institutions, which is as expected, people with relatively less power, or within more complex institutions like a national government, reported a similar level of personal empowerment. 

Most of the decision makers feel empowered both in terms of having purview in their role and in terms of being able to effect change in the organization itself. Even in the most tradition-bound organizations: for example a large religious congregation and a national senate, the individuals describe having the power to make major organizational changes. Notably in the face of external upheavals, which is an impetus for change long observed and recognized in social movement scholarship by the concept of “shock doctrine” (Klein, 2007) and in institutionalization and social sciences as “punctuated equilibria” (Aoki, 2000), wherein change can be more easily implemented when there is a sudden and external need. 

\subsection{Fostering institutional stability during times of change}

Consider Emerson (the Rabbi) who expressed how the onset of social isolation and quarantine drastically challenged their services and emphasized “the need for adaptability, creativity, and compassion in the face of crisis.” Highlighting the caregiving and community-driven nature of his organization, the pandemic compelled him into “a situation that required reimagining centuries of tradition within a matter of days. Religious services moved online, funerals were conducted with restrictions or via livestream, and neonatal circumcisions (brisses) required careful adaptation to ensure safety while honouring tradition.” Emerson’s experiences illustrate principles of sociological institutionalism, showing how the transition of religious services to online formats fosters the kinds of dependencies that sustain institutional stability over time (Greener and Powell, 2024), yet several questions arise regarding the mitigating role of technology within power relations: Is this a good thing that people can continue to attend religious services online? Does it offer connection from afar, or does it limit access to more affluent members of society? Or, does it represent something else entirely? How will AI change people’s power relations? What can be discerned about these realities, given the framework of the levers of power? We consider these questions in light of the technology’s double-edge, in which it affords the user power that can be used to various purposed. While technology has been shown to stabilize institutions, it also draws new boundaries. For instance, access to faith services during the global COVID-19 pandemic relied partially on broadband and also digital literacy. In this light, as AI becomes more prevalent, it is natural to question whether religious authority and truth will become more dependent on who controls the algorithms, rather than who resides at the pulpit or in the congregation. We must seriously consider technology as the mediation of not only religion but of community and relationships. 

\subsection{Influencing institutional change}

In terms of influencing institutional change, it is useful to note that the logic by which different decision makers operate does not automatically correlate with the logic one might associate with their sector. An executive from a large international tech company for example reports making some choices based on a logic of social good, and an academic is considering projects from a logic of cost-effectiveness. This indicates that personal relationships may be a key element for effective use of influence, even for groups who are approaching institutions, since those institutions are represented by people operating from varying logics, and may best be recruited into a coalition through a rationale that matches their personal preferences. One respondent, a senior political scientist, underlined this aspect in explaining that a most impactful event in their career had happened because a group of policymakers were given the time to read each other’s work before starting a policymaking process together. Other respondents explained how sudden change could be the seed of newly fit-to-purpose responses.

There are contradictions within the responses. For example, almost all the participants who responded to the questionnaire answered that stakeholders can reach out and get in touch with them, which contradicts the ground reality of disenfranchisement in society, which is to say that most of the respondents are inaccessible to the general public, due to the disposition of their institution, and structural constraints such as access to knowledge and social constructions that foster imposter syndrome. Even presuming an accessible institutional stance, individuals may have or perceive prohibitions in place that prevent them from engaging with people from other sectors, leading to contradictions between rules and practice, and a subsequent need for an institutional insider to find mechanisms to resolve the conflict or, from institutionalism, a “decoupling” (Boxenbaum, Jonsson (2017)). For example, our persona Jaime, a bureaucrat in a democracy, expresses their need to be securely anonymous, yet they subsequently responded in the affirmative to the question on whether they are accessible to the general public. This illustrates a potential decoupling tension in modern bureaucracies between traditional anonymity, designed in part for protection from political reprisals, and the contemporary demand for democratic public engagement. As stated by Jamie, “I am unable to answer most questions for ethical, political, safety and security reasons. Please do not attribute any information to myself, my office or department.” The work that dedicated public servants do to resolve decoupling challenges is one of the strongest ways for institutions to respond to new situations, in terms of effectiveness, speed, and staying power of new, re-coupled, practices.

\subsection{Hypotheses on the dynamics of the levers of power}

The cross-referencing of questionnaire responses and the documentary examination of respondent's professional body of work provides the means to offer insight on the status of different levers of power, in terms of how individuals can access and lend their personal power to these levers, for the purpose of steering the activities of institutions, organizations, and civil society advocacy groups on the questions of governing AI in society. This table provides a list of “levers of the power” and the status of each lever in terms of the dynamics of how it is being used to affect change in the institutional field of AI, as at this writing.

\begin{longtable}{@{}p{5cm}p{10cm}@{}}
\toprule
\textbf{Lever of Power} & \textbf{Status} \\ \midrule
Logics & Market logic, social justice logic, technosaviourism \\
Relative elaboration of institutional infrastructure & Emergent and translated \\
Governance & Incomplete coverage \\
Collective interest organizations & Few and far between \\
Regulators & Lacking in purview \\
Informal governance bodies & Inconclusive \\
Field configuring events & Present, supporting multiple logics \\
Status differentiators & Money, resources, and authority \\
Organizational models or templates & Tech industry mixed standard \\
Categories / Labels & High potential for impact \\
Norms & In flux; accessible \\
Relational channels & Formal, in transformation \\
Idea mobility & Low integration, high variance \\
\bottomrule
\end{longtable}

\noindent
\textit{Note: For a full description of terms and the details of the analysis, please request more information from the authors.} 
 
From the responses and from studying the work of the respondents, several dynamics can be hypothesized and tested: 

\begin{itemize}
    \item H1: Formal methods are outpaced, with increasing leverage for informal methods.
    \item H2: Institutional dynamics are dominated by business marketing and community disengagement.
    \item H3: Formal and informal power mechanisms are being strategically addressed by big tech.
    \item H4: Informal influence remains elitist and siloed.
    \item H5: Collective action is growing rapidly through large-scale initiatives.
\end{itemize}

\section{Limitations}

Power is not always benevolent, and readers should take special care that our recommendations are being used to build coalitions to represent community needs, not the contrary.

The findings may or may not be generalizable. We did not seek statistical correlations or completeness. In keeping with ethnographic practice, this research provides rich detail on the context and circumstances of individuals and infers potential cultural patterns from similarities and disparities in responses. This study spends considerable time focusing on each individual’s context and experiences in an attempt to ensure their unique stories are brought to life. While it might be easy to attempt to generalize the study's findings through a variety of “isms,” a reductionist perspective may lessen each individual's experiences and ignore what makes their positions and communities special and different. From this perspective, the uniqueness of the participants' world-views, and the inherent personal value of their experiences should be both recognized and celebrated. Personal power is inseparable from personal context.

Though they came from several continents and multiple cultures, all of the respondents are currently working in North America or Europe. Future studies may expand the questionnaire as a standardized survey to larger groups of decision makers, but this remains to be determined, as decision makers are notoriously hard to get into survey-responding groups.

\section{Conclusion}

The importance of understanding power dynamics and the pressures on institutional change is underlined by the scientific director of a major AI institution, who responded to our question about their biggest concerns on AI. They said: “We do not know yet how we would control AI that could be smarter than us, yet we collectively race forward, taking an existential risk for humanity's future.” We consider this to be an excellent and poignant argument, reminiscent of how we have seen technology created and manipulated in history. Consider nuclear fission, which was first designed to be a cleaner power, then was taken and used to destroy Nagasaki and Hiroshima. We see people in power using and manipulating technology for their own gains throughout history. How will AI be used and manipulated moving forward? We assert that we need decision-makers to be enacting policy for the greater-good. Such a monumental task has to be managed by policymakers within institutions and their counterparts in civil society. This study provides insight on the means for these institutional change-makers to personally engage with AI governance.

Further research on the dynamics of institutionalization in the field of AI and of the questionnaire responses is ongoing. Researchers interested in collaborating are invited to get in touch.

\bibliographystyle{apalike}   
\nocite{*}
\bibliography{references}     

@incollection{Aoki2000,
  author    = {Masahiko Aoki},
  title     = {Institutional evolution as punctuated equilibria},
  booktitle = {Institutions, Contracts and Organizations: Perspectives from New Institutional Economics},
  editor    = {Claude Ménard},
  year      = {2000},
  publisher = {Edward Elgar},
  address   = {Cheltenham},
  pages     = {11--33}
}

@incollection{Boxenbaum2017,
  author    = {Eva Boxenbaum and Stefan Jonsson},
  title     = {Isomorphism, diffusion and decoupling: Concept evolution and theoretical challenges},
  booktitle = {The Sage Handbook of Organizational Institutionalism},
  editor    = {Royston Greenwood and Christine Oliver and Thomas B. Lawrence and Renate E. Meyer},
  year      = {2017},
  publisher = {Sage},
  address   = {Thousand Oaks, CA},
  pages     = {77--101}
}

@incollection{Breeze2018,
  author    = {Maddie Breeze},
  title     = {Imposter syndrome as a public feeling},
  booktitle = {Feeling Academic in the Neoliberal University: Feminist Flights, Fights and Failures},
  editor    = {Yvette Taylor and Kinneret Lahad},
  year      = {2018},
  publisher = {Springer},
  address   = {Cham},
  pages     = {191--219}
}

@incollection{Greener2024,
  author    = {Ian Greener and Martin Powell},
  title     = {Historical institutionalism and social policy},
  booktitle = {Handbook on the Political Economy of Social Policy},
  editor    = {Bent Greve and Amílcar Moreira and Minna van Gerven},
  year      = {2024},
  publisher = {Edward Elgar},
  address   = {Cheltenham},
  pages     = {14--24}
}

@incollection{Hinings2017,
  author    = {Christopher Robin Hinings and Danielle Logue and Charlene Zietsma},
  title     = {Fields, institutional infrastructure and governance},
  booktitle = {The Sage Handbook of Organizational Institutionalism},
  editor    = {Royston Greenwood and Christine Oliver and Thomas B. Lawrence and Renate E. Meyer},
  year      = {2017},
  publisher = {Sage},
  address   = {Thousand Oaks, CA},
  pages     = {163--189}
}

@article{Siddiqui2024,
  author    = {Zaha Kamran Siddiqui},
  title     = {Navigating imposter phenomenon: A collective journey towards empowerment},
  journal   = {Clinical Teacher},
  year      = {2024},
  volume    = {21},
  number    = {5},
  pages     = {e13783},
  doi       = {10.1111/tct.13783}
}

@article{Kandikatla2025,
  author    = {Laxmiraju Kandikatla and Branislav Radeljić},
  title     = {AI and human oversight: A risk-based framework for alignment},
  journal   = {arXiv},
  year      = {2025},
  month     = {October},
  doi       = {10.48550/arXiv.2510.09090}
}

@inproceedings{Khan2025,
  author    = {Rubaina Khan and Tammy Mackenzie and Sreyoshi Bhaduri and others},
  title     = {Whole-person education for AI engineers},
  booktitle = {Proceedings of the Canadian Engineering Education Association},
  year      = {2025},
  doi       = {10.24908/pceea.2025.19600}
}

@book{Klein2007,
  author    = {Naomi Klein},
  title     = {The Shock Doctrine: The Rise of Disaster Capitalism},
  year      = {2007},
  publisher = {Picador},
  address   = {New York}
}

@inproceedings{Mackenzie2024,
  author    = {Tammy Mackenzie and Leslie Salgado and Sreyoshi Bhaduri and others},
  title     = {Beyond the algorithm: Empowering AI practitioners through liberal education},
  booktitle = {2024 ASEE Annual Conference \& Exposition},
  year      = {2024}
}

@article{Mackenzie2025,
  author    = {Tammy Mackenzie and Branislav Radeljić and Olivia Heslinga},
  title     = {Easy to read, easier to write: The politics of AI in consultancy trade research},
  journal   = {Cogent Social Sciences},
  year      = {2025},
  volume    = {11},
  number    = {1},
  pages     = {2470368},
  doi       = {10.1080/23311886.2025.2470368}
}

@article{Malmqvist2019,
  author    = {Johan Malmqvist and Kristina Hellberg and Gunvie Möllås and others},
  title     = {Conducting the pilot study: A neglected part of the research process? Methodological findings supporting the importance of piloting in qualitative research studies},
  journal   = {International Journal of Qualitative Methods},
  year      = {2019},
  volume    = {18},
  doi       = {10.1177/1609406919878341}
}

@misc{Mila,
  author = {{Mila Quebec AI Institute}},
  title = {Summer School in Responsible AI and Human Rights},
  howpublished = {\url{https://mila.quebec/en/continuing-education/summer-school-in-responsible-ai-and-human-rights}},
  note = {Accessed: 02-Nov-2025},
  year = {n.d.}
}

@misc{New2020,
  author    = {Mark New and Jesse Maria-Kinney},
  title     = {Research for Impact Coursera MOOC},
  year      = {2020},
  note      = {Retrieved from \url{https://www.coursera.org/learn/research-for-impact}}
}

@article{Olsen2009,
  author    = {Johan P. Olsen},
  title     = {Change and continuity: An institutional approach to institutions of democratic government},
  journal   = {European Political Science Review},
  year      = {2009},
  volume    = {1},
  number    = {1},
  pages     = {3--32},
  doi       = {10.1017/S1755773909000022}
}

@article{Powell1983,
  author    = {Walter W. Powell and Paul J. DiMaggio},
  title     = {The iron cage revisited: Institutional isomorphism and collective rationality in organizational fields},
  journal   = {American Sociological Review},
  year      = {1983},
  volume    = {48},
  number    = {2},
  pages     = {147--160},
  url       = {http://www.jstor.org/stable/2095101?origin=JSTOR-pdf}
}

\appendix
\section*{Annex: Sample Questionnaire (Academic Template)}

Hello, (participant/academia name),

\subsection*{Introduction and Context}

This is a questionnaire to help understand how AI is developing as a field. As a decision maker in our society, you will have decisions to make that concern the governance of AI. The purpose of this study is to inform the design of a set of control measures (indicators and contingencies) for existing and upcoming AI situations, specifically to do with your role in society and in your work.

To that end, this questionnaire has been made specifically for you. It should take about 15 minutes. It explores the dimensions of governance in which you may engage. The goal of the questions is to understand aspects of governance in your field. Please, answer each question as completely as possible.

If you have any questions, comments, or concerns, please use the comment / other fields or reply by email.

\subsection*{Ethics}

You may opt-out of any question. There is no word limit on responses. If you run out of space, you can email the lead author, Tammy Mackenzie, at \texttt{tmackenziemontreal@gmail.com}. There is an opportunity to comment on each of the 5 sections of the questionnaire.

\subsection*{Privacy}

You may choose to have your responses anonymized, which is to say your role but not your name will be included with your responses. For example, if anonymous, you might be referred to as ``a PhD student at a university (in your field), (your field of study).'' If not anonymous, with ``Your name, your department or field, YourU/Org, your field of study.'' 

The data is being gathered in this Google Form, and will be kept long-term, and will be backed up with Microsoft encryption on the PC of the lead researcher, in Canada, and backed up to a back-up service, also in Canada. Your responses may be included in whole or in part in future studies or publications. You may request to know what data has been gathered about you.

No questions require you to disclose private information. Please do not provide confidential information.

\subsection*{GPT use}

We humans have written every word of this work. We asked GPT for advice throughout the process, and have from time to time taken it.

\textbf{YOU MAY OPT OUT OF ANY QUESTION EXCEPT THE FIRST TWO, ON THIS PAGE,} which is for your email address and your choice on anonymity. 

\subsection*{SECTION 1 of 5: ANONYMITY}
\begin{enumerate}
    \item Do you want your responses to be anonymous?\\
    $\square$ Anonymous \\
    $\square$ Not anonymous
    \item Comments / questions / concerns? 
\end{enumerate}

\subsection*{SECTION 2 of 5: DEMOGRAPHICS}
\begin{enumerate}
    \item Did We Get This Right? Does this accurately describe your role? (Description of their role and their institution). Please help us to correct any misconceptions:
    \item Please describe your cultural background in one or two sentences:
    \item Comments / questions / concerns? 
\end{enumerate}

\subsection*{SECTION 3 of 5: ABOUT YOUR INSTITUTION}
\begin{enumerate}
    \item On a scale of 1 to 5, how would you rate your institution in terms of maturity (how established and widely accepted it is):
    \item On a scale of 1 to 5, how would you rate your institution in terms of complexity (the interdependence and variety of different functions)?
    \item On a scale of 1 to 5, rate the role of tradition in your daily work:
    \item Would you describe your organization as hierarchical or horizontal?\\
    $\square$ Hierarchical \\
    $\square$ Horizontal \\
    $\square$ Mixed / depends
    \item Which of these best describes the organizational structure that supports your role on a day-to-day basis: \\
    $\square$ no organizational structure \\
    $\square$ we sometimes try new things \\
    $\square$ we have rules and norms that are sometimes changed \\
    $\square$ it's always been done this way
    \item Do you have a superior who can make major changes to your organization or industry? If so, can you speak with them informally? Can you go have a coffee or a walk together, for example?
    \item Do you participate in the elaboration of regulations in your institution?
    \item Are your interactions with your subordinates and superiors governed by a formal policy?
    \item If your answer was yes, can you share a link to this policy?
    \item A Typical Week: We'd like to understand your interactions with people inside and outside your institution. Please, describe or list some of the main activities and engagements you might have in a typical work week?
    \item Comments / questions / concerns? 
\end{enumerate}

\subsection*{SECTION 4 of 5: YOUR INTERACTIONS WITH SOCIETY}
\begin{enumerate}
    \item Do you interact with collective interest organizations?
    \item Which collective interest organizations do you interact with within your work?
    \item On a scale of 1 to 5, rate the level to which your actions are scrutinized by collective interest organizations:
    \item Different from collective interest organizations, informal governance bodies can include professional orders, clubs, social and professional commissions, networks, cultural, lifestyle, or theological groups, ad-hoc gatherings that occur regularly, etc. On a scale of 1 to 5, what influence do informal governance bodies have on your work? 
    \item Which informal governance bodies do you interact with within your work?
    \item Do you answer to an ombudsman or professional order? \\
    $\square$ Ombudsman \\
    $\square$ Professional order \\
    $\square$ No formal oversight
    \item If yes, please provide a link to any professional codes of conduct or other relevant regulation:
    \item Can stakeholders get in touch with you for a one-on-one talk?\\
    $\square$ Yes \\
    $\square$ No \\
    $\square$ Some of them yes, some of them no
    \item How can stakeholders communicate with you?
    \item Can you give an example of when your institution responded very well to an event, situation, or otherwise important moment? Why was it good?
    \item Can you give an example of when your institution responded very poorly to an event, situation, or otherwise important moment? Why was it poor?
    \item Comments / questions / concerns? 
\end{enumerate}

\subsection*{SECTION 5 of 5: YOUR ROLE IN YOUR ORGANIZATION}
\begin{enumerate}
    \item Do people generally understand what you do in your daily work? \\
    $\square$ Yes \\
    $\square$ No \\
    $\square$ not sure
    \item You need to communicate a new idea to the people in your organization. On a scale of 1 to 5, how many people will you reach? \\
    1- a few people / my department \\
    2- most people in my organization \\
    3- everyone in my organization \\
    4- all of our stakeholders \\
    5- the general public
    \item You are considering a complex issue at work. Which aspects of the problem concern you, in order of priority? (rate any that apply - please only use 1 rank per aspect.)
    \begin{itemize}
        \item Ethical/cultural
        \item Political
        \item Technological
        \item Economic
        \item Organizational
        \item Other
    \end{itemize}
    \item If you answered "other", you may provide more information here: 
    \item On a scale of 1 to 5, rate your personal level of influence on governance in your institution:
    \item Scenario A: There is an environmental problem in your field. Which solution would you promote?\\
    $\square$ Support research on a promising technical solution \\
    $\square$ Make a training program for specialists
    \item Scenario B: There is a social problem in your field. Which solution would you promote?\\
    $\square$ Create a scholarship program \\
    $\square$ Create an entrepreneurship program for graduates
    \item Scenario C: A market vital to your organization is collapsing. Which solution would you promote?\\
    $\square$ Organize a conference of industry and academic specialists \\
    $\square$ Organize a tech upskilling program
    \item An Impactful Event: Briefly describe a major event that changed how you see or do your work. You can consider questions such as: Who participated? What was the intent or reason for the event? What impact did it have on you? On your field? (Please include links if possible.)
    \item Did you “rise through the ranks"?
    \item How long have you been in your position? 
    \item Your Superhero Name: We sometimes may not realize how important we can be to other people in society. In this question, we hope to understand what your highest aspiration is for your role in society. On your best day, how would you describe it? This is one way that may seem silly but that helps us understand, in institutional terms, the norms and labels that others can recognize as archetypes. Please, imagine you’ve been asked to collaborate on a comic book about your work, and you must give yourself a superhero title. Examples: Super-Strong People Defender! Maker of Amazing Solutions! 
    \item Do you have some concerns about AI in your work? Please list or describe. 
    \item Comments / questions / concerns? 
\end{enumerate}

\end{document}